\documentclass[twocolumn,showpacs,preprintnumbers,amsmath,amssymb,prl]{revtex4}

\usepackage{graphicx}
\usepackage{dcolumn}
\usepackage{bm}

\begin{document}


\title{ {\em Ab initio} mechanical response: internal friction and structure of divacancies in silicon.}

\author{H. \"Ust\"unel}
\affiliation{Department of Physics, Cornell University, Ithaca, NY 14853}

\author{D. Roundy}
\affiliation{Department of Physics, Cornell University, Ithaca, NY 14853}

\author{T.A. Arias}
\affiliation{Department of Physics, Cornell University, Ithaca, NY 14853}


\begin{abstract}
  This letter introduces {\em ab initio} study of the full
  activation-volume tensor of crystalline defects as a means to make
  contact with mechanical response experiments.  We present a
  theoretical framework for prediction of the internal friction
  associated with divacancy defects and give the first {\em ab initio}
  value for this quantity in silicon.  Finally, making connection with
  defect alignment studies, we give the first unambiguous resolution
  of the debate surrounding {\em ab initio} verification of the
  ground-state structure of the defect.
\end{abstract}

\maketitle

Among the most powerful applications of first principles {\em ab
  initio} approaches in condensed matter systems is the interpretation
of experimental signatures from defects.  The extremely efficient,
albeit approximate, functionals available for density-functional
theory have given this approach wide application in conjunction with
experimental probes such as scanning tunneling microscopy, electron
paramagnetic resonance, electron-nuclear double resonance, and nuclear
magnetic
resonance\cite{Shaikhutdinov,McMahon,Carroll,Gabor,Goss,Ogut,Overhof,Weihrich,Mauri,Zhang,Mauri2}.
Macroscopic mechanical response experiments also have proved an
extremely valuable tool in the study of defects in solid-state
systems\cite{Pohl,Cannelli,Snead,Liu,Laszig}, providing key
information on such issues as defect symmetries and concentrations.
However, mechanical response studies remain largely ignored by the
{\em ab initio} community to date.

Despite their successes, the aforementioned density-functional studies
suffer a fundamental flaw: no underlying theorem ensures that
density-functional theory provides the energy- or
angular-momentum- resolved densities probed in the experiments involved in
the application.
Beyond this matter of principle, such quantities are not among those
which available approximate functionals predict most reliably.  This
letter notes that the most reliable quantities which density-functional
theory predicts (bond lengths, bond angles, and lattice parameters) relate
directly to the key coupling parameter in macroscopic mechanical
response experiments, the full activation-volume {\em tensor}.  We
thus propose {\em ab initio} study of mechanical response functions
associated with this tensor as a powerful and particularly reliable
new tool in the study of defects in condensed matter systems.

One such response function, internal friction, is a topic of current
interest both experimentally and
theoretically\cite{Cleland,Lifshitz1,Lifshitz2,Lidija,Evoy,Hutchinson}.
Below we develop the theory of internal friction as applied to
divacancy defects to provide the first parameter-free {\em ab initio}
determination of friction from a point defect, in this case the singly
negatively charged divacancy in silicon (Si-$V_2^-$).  Although the
structure of this defect has been inferred through a
combination of experimental signatures and symmetry of electronic
states\cite{WC}, {\em ab initio} work to confirm the structure has
resulted in an ongoing debate\cite{Ogut,Saito,WReply,SReply} which 
has arisen ultimately from a focus on delicate energy differences at or
beyond the limits of accuracy of current density functionals.  Here,
we demonstrate the power of working with mechanical response functions
by providing a clear signature which resolves the debate.

{\em Mechanical response of defects ---} Within linear response, the
stress-dependent part of the energy of a point defect $\Delta
E$ has the form
\begin{equation} \label{eq:DE}
\Delta E=-\Lambda_{ij} \sigma_{ij},
\end{equation}
where we employ repeated index summation notation here and throughout
this work and where $\sigma$ and $\Lambda$ are, respectively, the
externally applied stress tensor and the {\em defect activation-volume
  tensor}.

The activation-volume tensor $\Lambda$ is accessible in terms of
equilibrium supercell lattice constants, which are among the
quantities most reliably determined within density-functional theory.
This connection comes directly from the principle that the strain $u$
induced in a crystal containing a concentration $n$ of defects with
activation volume tensor $\Lambda$ is $u_{ij}=n \Lambda_{ij}$, a
result of minimizing the sum of the bulk and defect elastic energies.
Accordingly, one may determine $\Lambda$ simply by creating a
supercell containing a single defect and relaxing the defect structure
along with the supercell lattice vectors.

Under applied stress, defects with lower symmetry than the host
crystal tend to reorient so as to minimize the energy.  Experimental
stress-alignment studies, which observe the relative thermalized
populations of different orientations of specific defect types as a
function of applied stress, thus allow direct access to certain linear
combinations of $\Lambda_{ij}$ for each type~\cite{WC}.  Such
thermalization of defect populations under {\em time-varying} external
stress $\sigma_{ij}(t)$ provide a mechanism for internal friction,
dissipation of mechanical energy throughout the bulk of a material.
We review this process here in some depth because, although our
overall logic is the same, the final result for the divacancy differs
from the oft-quoted result \cite{book}, which assumes that all defect
orientations relax among each other at equal rates and hence does
not apply to divacancies.

From (\ref{eq:DE}), the total energy per unit volume stored at
time $t$ among all defect orientations $m$ of a specific defect type is
\begin{equation}
\Delta E(t) = -n \sum_m P^m(t) \Lambda^m_{ij} \sigma_{ij}(t),
\end{equation}
where $n$ is the total number density of defects of this type, and
$P^m$ and $\Lambda^m_{ij}$ are the probability and
activation volume tensor associated with each orientation.
Dissipation results from energy lost irreversibly to the heat bath
through transitions among defect orientations and thus occurs at the
rate
\begin{equation} \label{eq:loss}
\frac{dE}{dt}= n \sum_m \frac{d P^m}{d t} \Lambda^m_{ij} \sigma_{ij}(t).
\end{equation}
Finally, stresses ultimately drive the transitions through the
master equation
\begin{eqnarray} \nonumber
\frac{dP^m}{dt} & = & \sum_{m'}{
\nu_{mm'} \left(1+\beta  \Lambda^{m'}_{ij}
  \sigma_{ij}(t)\right)P^{m'}(t)} \\
&& -
\nu_{mm'} \left(1+\beta  \Lambda^m_{ij}
  \sigma_{ij}(t)\right)P^m(t), \label{eq:ME}
\end{eqnarray}
where we take transitions from orientation $m'$ to $m$ to be
thermally activated
with rate $\nu_{mm'}$ in the absence of external stress and with stress
dependencies which we have linearized.  Solving (\ref{eq:ME}) and
substituting the result into (\ref{eq:loss}) then gives the final
dissipation rate.

\def\re{{\mbox{Re}}}
\def\im{{\mbox{Im}}}
\newcommand{\omegaovernug}{\frac{\omega}{\nu g}}
\newcommand{\sigij}{\tilde \sigma_{ij}^*}

Under the not uncommon special case underlying the result \cite{book},
in which the zero-stress transition rates among different orientations
of a defect type are equal by symmetry, $\nu_{mm'}=\nu$, the master
equation may be solved analytically, resulting in a defect
contribution to the inverse quality factor $Q^{-1}$, the
fraction of energy lost per radian of oscillation phase, of
\begin{align} \label{eq:Qinv}
  Q^{-1} &= n \beta
  \frac{\omegaovernug}
  {1+\left(\omegaovernug\right)^2}
  \frac{\sigij L_{ij;kl} \tilde \sigma_{kl}}
  {\sigij S_{ij;kl} \tilde \sigma_{kl}}\textrm{,} \\
  L_{ij;kl} &\equiv \frac1g \sum_m
  \Delta \Lambda_{ij}^m \, \Delta \Lambda_{kl}^m. \nonumber
\end{align}
Here, $\sigma_{ij}(t) \equiv \re\left(\tilde \sigma_{ij}
  \exp\left(i\omega t\right)\right)$, $\Delta \Lambda_{ij}^m \equiv
\Lambda^m_{ij} - \frac1g \sum_{m'} \Lambda^{m'}_{ij}$, $S_{ij;kl}$ is the
standard elastic compliance four-tensor, and $L_{ij;kl}$ is an {\em
  anelastic four-tensor} sharing the symmetry of the full set of
defect orientations.

In a system with more than one type of defect, the inverse quality
factors for each type add.  If the transition rate $\nu_s$ associated
with each type $s$ has the same value $\nu$, then $Q^{-1}$ can be
written as (\ref{eq:Qinv}) with
\begin{equation} \label{eq:Ltot}
  L^{tot}_{ij;kl} \equiv \sum_s \frac{n_s}{n} L^s_{ij;kl}
\end{equation}
replacing $L_{ij;kl}$, where $n_s$ is the number density for each type of
defect so that $\sum_s n_s \equiv n$.  Below we show how to apply
Eqs.~(\ref{eq:Qinv},\ref{eq:Ltot}) to find the correct result for
divacancies.

Finally, we note here that there is a simple, exact relationship
between the anelastic four-tensor and the results from experimental
studies of the inverse quality factor of a mechanical oscillator as a
function of temperature, which generally show a peak when the
thermally activated, and thus temperature-dependent, transition rate
$\nu(T)$ corresponds to the oscillator frequency\cite{Pohl}.  In
particular, from (\ref{eq:Qinv},\ref{eq:Ltot}) we have that the
experimentally accessible quantity $k_B T Q^{-1}$, where $k_B$ is
Boltzmann's constant, has a maximum at precisely $\nu(T)=\omega/g$
with value
\begin{equation} \label{eq:TQinv}
\max\left(k_B T Q^{-1}\right) = \frac{n}{2} \frac{\tilde
  \sigma_{ij}^* L^{tot}_{ij;kl} \tilde \sigma_{kl}}{\tilde \sigma_{ij}^* S_{ij;kl}
  \tilde \sigma_{kl}}.
\end{equation}
{\em Ab initio} determination of $L^{tot}_{ij;kl}$ thus gives a
parameter-free relationship between the maximum temperature-internal
friction product and the defect concentration, thereby allowing mechanical
response experiments to provide a direct, parameter-free measure of absolute
defect concentrations for the first time.

{\em Application to divacancy in silicon ---} The importance of the
divacancy in silicon has led to its extensive study, both theoretically and
experimentally\cite{Song,WC,Saito,Pesola,Ogut,Cheng,Svensson}.  In contrast
to single vacancies, which are quite mobile and thus anneal readily,
divacancies have low mobility and are among the most common stable defects
in silicon at room temperature.  The defect has four charge states, singly
positively charged, neutral, singly and doubly negatively charged, with the
singly charged defect (Si-$V_2^-$) playing an important role in carrier
recombination~\cite{Hallen}.

Since the publication of Watkins and Corbett's pioneering study \cite{WC},
the {\em ab initio} determination of the precise nature of the ground-state
structure of Si-$V_2^-$ has become the subject of
debate\cite{Ogut,Saito,WReply,SReply}.  The idealized defect has $D_{3d}$
symmetry along the $\left< 111 \right>$ axis connecting the sites of the
neighboring vacant atoms.  The defect also introduces partially filled
degenerate electronic states into the gap and thus undergoes a Jahn-Teller
distortion which ultimately lowers the symmetry to $C_{2h}$.  The debate
arises because two stable structures, termed {\em pairing} and {\em
resonant}, with very similar energies are consistent with the $C_{2h}$
symmetry of the defect.

\begin{figure}
\begin{center}
\scalebox{0.3}{\includegraphics{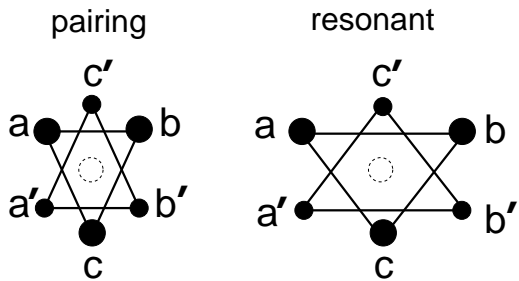}}
\end{center}
\caption{Projection of divacancy structure along $\langle 111\rangle$ symmetry axis:
  missing atoms (dashed circles), atoms in plane closer to observer
  (larger solid circles), atoms in plane further from observer
  (smaller solid circles).}
\label{fig:config}
\end{figure}

Figure~\ref{fig:config} shows the two competing ground-state
structures as projected along the $\left< 111 \right>$ {\em defect
  axis} connecting the sites of the two vacant atoms.  The pairing
configuration breaks symmetry by moving two pairs of atoms ($ab$ and
$a'b'$ in the figure) toward each other along a $\left< 110 \right>$
{\em reconstruction axis} to form stronger bonds at the expense of
strain energy.  In the resonant structure, the same pairs of atoms
move away from each other to form a less favorable bonding
configuration at a nearly correspondingly lower cost in strain.
Whenever this work requires a specific coordinate system, it shall be
such that the {\em defect axis} is $[111]$ and the {\em bonding axis}
is $[1\bar10]$.

The literature reports a number of density-functional values for the energy
difference $\Delta E \equiv E_{{\mathrm pair}}-E_{{\mathrm res}}$ between
the two competing configurations: $0.0024$~eV\cite{Saito}, $\sim
0$~eV\cite{Pesola}, $-0.12$~eV\cite{Ogut}, $-0.18$~eV (present work).  The
fact that these differences are all quite small and of the order of the
uncertainties due to the approximate density functionals which these works
employ (all other computational uncertainties notwithstanding!) underscores
the difficulty of using total energies to resolve the ground state structure
of the Si-$V_2^-$ defect and indicates that previous studies are
inconclusive.  We propose instead to use the considerably different
activation volume tensors of the competing reconstructions as a more
appropriate signature to confirm the ground state.

Experimentally, Watkins and Corbett~\cite{WC} explored the activation volume
tensor of the Si-$V_2^-$ defect in depth using electron spin resonance to
study thermal alignment of defect subpopulations under external $\langle
110\rangle$ stresses.  These experiments, as do also internal friction
experiments, take place under conditions where $\langle 110 \rangle$
reconstruction axes have time to thermalize while $\langle 111 \rangle$
defect axes do not.  Such $\langle 110\rangle$ stresses split divacancies
into two classes, $\alpha$ and $\beta$, according to the orientation of
the defect axis, with $\alpha$ corresponding to the defect axis being
perpendicular to the stress.  Within each of these classes, there is a
further splitting of the energy into two distinct values, for a total of
four energetically different states~\cite{IR}.

As thermalization occurs only among choices of reconstruction axis,
the quantities which stress-alignment experiments actually access are
the energy splittings within each class, $\Delta E_\alpha$ and $\Delta
E_\beta$, respectively, each of which relate directly to certain
linear combinations of components the defect activation-volume tensor,
\begin{align}
\Delta E_\alpha & = -\frac{\sigma}{2}
\left( \Lambda_{11}+2\Lambda_{12}-2\Lambda_{13}-\Lambda_{33} \right)
\label{alpha} \equiv -\sigma \Delta\Lambda_\alpha\\
\Delta E_\beta &= -\frac{\sigma}{2} \left(
\Lambda_{11}-2\Lambda_{12}+2\Lambda_{13}-\Lambda_{33}
\right)\equiv -\sigma \Delta\Lambda_\beta,
\nonumber 
\end{align}
where $\sigma$ is the magnitude of the external $\left<110\right>$
stress and $\Lambda_{ij}$ are the Cartesian components of the
activation volume tensor in the cubic coordinate system defined above.

{\em Calculations and results ---} The {\em ab initio} electronic
structure calculations below employ the total-energy plane-wave
density-functional pseudopotential approach~\cite{RMP} within the local
spin-density approximation (LSDA) using a pseudopotential of the
Kleinman-Bylander form~\cite{KB} with $p$ and $d$ non-local corrections.
The calculations expand the Kohn-Sham orbitals in a plane-wave basis
set with a cutoff energy of 6~Hartrees within a cubic sixty-four atom
supercell, sampling the Brillioun zone at eight k-points, reduced to
four by time reversal symmetry.  Finally, we employ the analytically
continued functional approach~\cite{ACprl} to minimize the Kohn-Sham
energy with respect to the electronic degrees of freedom.

To determine $\Lambda$, we relax a supercell containing a single
defect and compute the relaxed strain and thus $\Lambda$.  In
principle, this can be done in a single joint relaxation with the
internal atomic coordinates.  For the present study, however, we
followed the nearly equivalent approach of minimizing the internal
coordinates while minimizing separately along each of the six
independent components of supercell strain while including the Poisson
ratios of the supercell when appropriate.

{\"O}${\breve {\text{g}}}${\"u}t and Chelikowsky\cite{Ogut} emphasize the need
to tailor supercell shape to accommodate the relaxation pattern of
$V_2^-$ to obtain accurate results for defect energy differences.  To
explore supercell-size effects on the extraction of the
activation-volume tensor, we carried out a convergence study of
$\Delta \Lambda_\alpha$ and $\Delta \Lambda_\beta$ from (\ref{alpha})
employing the environment-dependent interatomic potential (EDIP) for
silicon\cite{EDIP} for cubic supercells of lattice constant from $2
a_0$ through $6 a_0$, where $a_0$ is the lattice constant of the
``primitive'' eight-atom cubic cell.  Over this range of cell sizes,
we observe a total change of only $12.5\%$ ($6.3\%$) in $\Delta
\Lambda_\alpha$ ($\Delta \Lambda_\beta$), with 70\% (86\%) of the
change occurring between $2 a_0$ and $3 a_0$.  We thus conclude that
$\Lambda$ is not a particularly sensitive function of cell size and
that a supercell of 64 atoms suffices to give {\em ab initio} values
with an uncertainty on the order of 10\%.

\begin{figure}
\begin{center}
\scalebox{0.3}{\includegraphics{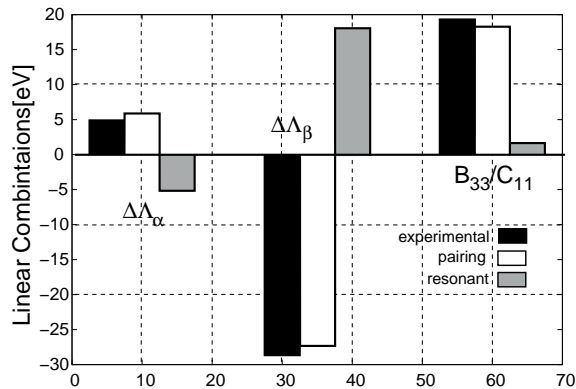}}
\end{center}
\caption{DFT results and comparison to experiment of $\Delta
\Lambda_\alpha$, $\Delta\Lambda_\beta$ and $B_{33}$. For each data set
the black bar corresponds to the experimental data, the white bar
to our DFT results for the pairing configuration and
the gray to those for the resonant configuration.  $B_{33}$ has
been scaled by $C_{11}$ for display purposes.}
\label{fig:barfig}
\end{figure}

Table~\ref{tab:Lambda} summarizes our {\em ab initio} results for the
activation volume tensors of the two candidate defect structures, and
Figure~\ref{fig:barfig} compares our predictions directly with the
experimentally available linear combinations, $\Delta \Lambda_\alpha$ and
$\Delta \Lambda_\beta$ \cite{WC}.  Our predictions for the resonant
configuration are clearly inconsistent with the measurements, whereas our
results for the pairing configuration show good agreement with errors (+20\%
and -6\% for $\Delta \Lambda_\alpha$ and $\Delta \Lambda_\beta$,
respectively) consistent with supercell-convergence uncertainty we estimate
above.  The figure also compares our {\em ab initio} prediction of another
linear combination of components of the activation volume tensor, $B_{33}
\equiv C_{33;ij} \Lambda_{ij}$ where $C_{ij;kl}$ is the elastic constant
four-tensor, with an {\em estimate} of this quantity from \cite{WC}.

\begin{table}
  \centering
  \hspace{-0.1in}
  \begin{tabular}{cc}
  \begin{tabular}{ccc}
    $\Lambda_\mathrm{pair}$ (\AA$^3$) && $\Lambda_\mathrm{res}$ (\AA$^3$) \\ \hline
$\left[\begin{array}{ccc}
 -11  &  18 &  1.5 \\
  18  & -11 &  1.5 \\
  1.5 & 1.5 & 10
\end{array}
\right]$ &&
$\left[\begin{array}{ccc}
 0.6  & -1.5 & 10 \\
 -1.5 & 0.6 &  10 \\
  10 &   10 & -13
\end{array}
\right]$
  \end{tabular}  &
\begin{tabular}{ccc}
$L^{tot} (\text{\AA}^6)$ & $pair$ & $res$  \\ \hline
$L_{11}$  & 100 &  38 \\
$L_{12}$  & -50 & -19 \\
$L_{44}$  & 293 & 279
\end{tabular} \\
\end{tabular}

\caption{Activation-volume and anelastic tensors for competing ground-state
  structures of singly negatively charged divacancy in silicon}
  \label{tab:Lambda}
\end{table}

Turning finally to the internal friction, because the defect axes do not
thermalize under typical experimental conditions, each choice of defect axis
must be treated as a separate \emph{type} of defect $s$ in (\ref{eq:Ltot}).
Under normal sample preparation, all four distinct choices of $\langle 111
\rangle$ defect axes will occur with equal probabilities, $n_s = n/4$,
resulting in an $L^{tot}$ with cubic symmetry. The thermalizing
reconstruction axes then constitute the orientations $m$ within each type
and have sufficient symmetry to ensure the result (\ref{eq:Qinv}).
Table~\ref{tab:Lambda} gives the three unique values of the resulting cubic
anelastic four-tensor in contracted notation.  This is the first {\em ab
initio} prediction for the components of the anelastic tensor for a defect,
a quantity of current research interest, particularly in the optimization of
micro- and nano- electromechanical devices.  (See, for example
\cite{Pohl,Lidija}.)  Our value of $L^{tot}$ indicates that torsional (pure
shear) oscillators will experience a maximum in divacancy-mediated loss in
the range $5.5\times10^{-28} \mathrm{cm}^3\mathrm{K} < \max(TQ^{-1})/n <
8.5\times10^{-28} \mathrm{cm}^3\mathrm{K}$, with the precise value depending
on the crystallographic orientation of the device through (\ref{eq:Qinv}).

In conclusion, this letter introduces {\em ab initio} study of the full
activation-volume tensor of crystalline defects as a quantity which current
approximate density-functionals give accurately and which is of direct use
in making contact with mechanical response experiments, including both
stress-alignment studies and measurements of macroscopic internal friction.
Illustrating the power of the approach, this letter gives the first
unambiguous {\em ab initio} verification of the nature of the ground state
of the singly negatively charged divacancy in silicon and the first
parameter-free theoretical calculation of the peak internal friction
associated with a point defect.  This latter quantity then forms the basis
for a straightforward method for determining defect concentrations via {\em
ab initio} interpretation of macroscopic mechanical response experiments.

Primary support of this work is through the MRSEC program of the NSF
(No. DMR-0079992).

\bibliographystyle{prsty} 
\bibliography{paper}

\end{document}